\documentstyle[twoside,psfig]{article}

\catcode`\@=11
\long\def\@makefntext#1{
\protect\noindent \hbox to 3.2pt {\hskip-.9pt  
$^{{\eightrm\@thefnmark}}$\hfil}#1\hfill}		

\def\@makefnmark{\hbox to 0pt{$^{\@thefnmark}$\hss}}	
	
\def\ps@myheadings{\let\@mkboth\@gobbletwo
\def\@oddhead{\hbox{}
\rightmark\hfil\eightrm\thepage}   
\def\@oddfoot{}\def\@evenhead{\eightrm\thepage\hfil
\leftmark\hbox{}}\def\@evenfoot{}
\def\sectionmark##1{}\def\subsectionmark##1{}}

\oddsidemargin=\evensidemargin
\newcounter{sectionc}\newcounter{subsectionc}\newcounter{subsubsectionc}
\renewcommand{\section}[1] {\vspace{12pt}\addtocounter{sectionc}{1} 
\setcounter{subsectionc}{0}\setcounter{subsubsectionc}{0}\noindent 
	{\tenbf\thesectionc. #1}\par\vspace{5pt}}
\renewcommand{\subsection}[1] {\vspace{12pt}\addtocounter{subsectionc}{1} 
	\setcounter{subsubsectionc}{0}\noindent 
	{\bf\thesectionc.\thesubsectionc. {\kern1pt \bfit #1}}\par\vspace{5pt}}
\renewcommand{\subsubsection}[1] {\vspace{12pt}\addtocounter{subsubsectionc}{1}
	\noindent{\tenrm\thesectionc.\thesubsectionc.\thesubsubsectionc.
	{\kern1pt \tenit #1}}\par\vspace{5pt}}
\newcommand{\nonumsection}[1] {\vspace{12pt}\noindent{\tenbf #1}
	\par\vspace{5pt}}

\newcounter{appendixc}
\newcounter{subappendixc}[appendixc]
\newcounter{subsubappendixc}[subappendixc]
\renewcommand{\thesubappendixc}{\Alph{appendixc}.\arabic{subappendixc}}
\renewcommand{\thesubsubappendixc}
	{\Alph{appendixc}.\arabic{subappendixc}.\arabic{subsubappendixc}}

\renewcommand{\appendix}[1] {\vspace{12pt}
        \refstepcounter{appendixc}
        \setcounter{figure}{0}
        \setcounter{table}{0}
        \setcounter{lemma}{0}
        \setcounter{theorem}{0}
        \setcounter{corollary}{0}
        \setcounter{definition}{0}
        \setcounter{equation}{0}
        \renewcommand{\thefigure}{\Alph{appendixc}.\arabic{figure}}
        \renewcommand{\thetable}{\Alph{appendixc}.\arabic{table}}
        \renewcommand{\theappendixc}{\Alph{appendixc}}
        \renewcommand{\thelemma}{\Alph{appendixc}.\arabic{lemma}}
        \renewcommand{\thetheorem}{\Alph{appendixc}.\arabic{theorem}}
        \renewcommand{\thedefinition}{\Alph{appendixc}.\arabic{definition}}
        \renewcommand{\thecorollary}{\Alph{appendixc}.\arabic{corollary}}
        \renewcommand{\theequation}{\Alph{appendixc}.\arabic{equation}}
        \noindent{\tenbf Appendix \theappendixc #1}\par\vspace{5pt}}
\newcommand{\subappendix}[1] {\vspace{12pt}
        \refstepcounter{subappendixc}
        \noindent{\bf Appendix \thesubappendixc. {\kern1pt \bfit #1}}
	\par\vspace{5pt}}
\newcommand{\subsubappendix}[1] {\vspace{12pt}
        \refstepcounter{subsubappendixc}
        \noindent{\rm Appendix \thesubsubappendixc. {\kern1pt \tenit #1}}
	\par\vspace{5pt}}

\newcommand{\textlineskip}{\baselineskip=13pt}
\newcommand{\smalllineskip}{\baselineskip=10pt}

\def\eightcirc{
\begin{picture}(0,0)
\put(4.4,1.8){\circle{6.5}}
\end{picture}}
\def\eightcopyright{\eightcirc\kern2.7pt\hbox{\eightrm c}} 

\newcommand{\copyrightheading}[1]
	{\vspace*{-2.5cm}\smalllineskip{\flushleft
	{\footnotesize Modern Physics Letters A, #1}\\
	{\footnotesize $\eightcopyright$\, World Scientific Publishing
	 Company}\\
	 }}

\newcommand{\publisher}[2]{{\begin{center}\footnotesize\smalllineskip 
	Received #1\\
	Revised #2
	\end{center}
	}}

\def\abstracts#1#2#3{{
	\centering{\begin{minipage}{4.5in}\footnotesize\baselineskip=10pt
	\parindent=0pt #1\par 
	\parindent=15pt #2\par
	\parindent=15pt #3
	\end{minipage}}\par}} 

\renewenvironment{thebibliography}[1]
	{\frenchspacing
	 \ninerm\baselineskip=11pt
	 \begin{list}{\arabic{enumi}.}
        {\usecounter{enumi}\setlength{\parsep}{0pt}     
	 \setlength{\leftmargin 12.7pt}{\rightmargin 0pt} 
         \setlength{\itemsep}{0pt} \settowidth
	{\labelwidth}{#1.}\sloppy}}{\end{list}}

\newcounter{itemlistc}
\newcounter{romanlistc}
\newcounter{alphlistc}
\newcounter{arabiclistc}

\newcommand{\fcaption}[1]{
        \refstepcounter{figure}
        \setbox\@tempboxa = \hbox{\footnotesize Fig.~\thefigure. #1}
        \ifdim \wd\@tempboxa > 5in
           {\begin{center}
        \parbox{5in}{\footnotesize\smalllineskip Fig.~\thefigure. #1}
            \end{center}}
        \else
             {\begin{center}
             {\footnotesize Fig.~\thefigure. #1}
              \end{center}}
        \fi}

\newcommand{\tcaption}[1]{
        \refstepcounter{table}
        \setbox\@tempboxa = \hbox{\footnotesize Table~\thetable. #1}
        \ifdim \wd\@tempboxa > 5in
           {\begin{center}
        \parbox{5in}{\footnotesize\smalllineskip Table~\thetable. #1}
            \end{center}}
        \else
             {\begin{center}
             {\footnotesize Table~\thetable. #1}
              \end{center}}
        \fi}

\def\@citex[#1]#2{\if@filesw\immediate\write\@auxout
	{\string\citation{#2}}\fi
\def\@citea{}\@cite{\@for\@citeb:=#2\do
	{\@citea\def\@citea{,}\@ifundefined
	{b@\@citeb}{{\bf ?}\@warning
	{Citation `\@citeb' on page \thepage \space undefined}}
	{\csname b@\@citeb\endcsname}}}{#1}}

\newif\if@cghi
\def\cite{\@cghitrue\@ifnextchar [{\@tempswatrue
	\@citex}{\@tempswafalse\@citex[]}}
\def\citelow{\@cghifalse\@ifnextchar [{\@tempswatrue
	\@citex}{\@tempswafalse\@citex[]}}
\def\@cite#1#2{{$\null^{#1}$\if@tempswa\typeout
	{IJCGA warning: optional citation argument 
	ignored: `#2'} \fi}}

\def\pmb#1{\setbox0=\hbox{#1}
	\kern-.025em\copy0\kern-\wd0
	\kern.05em\copy0\kern-\wd0
	\kern-.025em\raise.0433em\box0}

\def\fnt#1#2{\footnotetext{\kern-.3em
	{$^{\mbox{\scriptsize #1}}$}{#2}}}

\def\fpage#1{\begingroup
\voffset=.3in
\thispagestyle{empty}\begin{table}[b]\centerline{\footnotesize #1}
	\end{table}\endgroup}

\def\runninghead#1#2{\pagestyle{myheadings}
\markboth{{\protect\footnotesize\it{\quad #1}}\hfill}
{\hfill{\protect\footnotesize\it{#2\quad}}}}
\font\tenrm=cmr10
\font\tenit=cmti10 
\font\tenbf=cmbx10
\font\bfit=cmbxti10 at 10pt
\font\ninerm=cmr9

\font\eightrm=cmr8

\def\qed{\hbox{${\vcenter{\vbox{			
   \hrule height 0.4pt\hbox{\vrule width 0.4pt height 6pt
   \kern5pt\vrule width 0.4pt}\hrule height 0.4pt}}}$}}

\def\ra{\rightarrow}

\def\be{\begin{equation}}
\def\ee{\end{equation}}
\def\ba{\begin{eqnarray}}
\def\ea{\end{eqnarray}}

\begin{document}
\setlength{\textheight}{7.7truein}  

\runninghead{
A Test of SU(15) at HERA Using the HELAS Program
Manuscripts $\ldots$}{
A Test of SU(15) at HERA Using the HELAS Program
Manuscripts $\ldots$}

\normalsize\textlineskip
\thispagestyle{empty}
\setcounter{page}{1}

\copyrightheading{}			

\vspace*{0.88truein}

\fpage{1}
\centerline{\bf
A Test of SU(15) at HERA Using The HELAS Program
}
\vspace*{0.37truein}
\centerline{\footnotesize
Sun Myong Kim
\footnote{
skim@hit.halla.ac.kr, mplsmn@hanmail.net
}}
\baselineskip=12pt
\centerline{\footnotesize\it
Department of Liberal Arts and Sciences, Halla University
}
\baselineskip=10pt
\centerline{\footnotesize\it
WonJu, Kangwondo 220-840, Korea
}

\vspace*{0.225truein}

\publisher{(received date)}{(revised date)}

\vspace*{0.21truein}
\abstracts{
A possible SU(15) process at HERA is investigated.
The process that we consider is $e^- P\ra \bar\nu_e \mu^- \mu^- +anything$
through the exchange of new heavy gauge bosons $X^-$ and $X^{--}$ which are
predicted in SU(15).  This process produces two easily observable like-sign
muons in the final state.
The cross section of this process is calculated
by using HELAS and VEGAS programs, and PDF-library functions.
The cross section turns out to be small to be observed in near future.
}{}{}



%
%
\vspace*{25pt}
\noindent

The standard model,
SU(3)$\otimes$SU(2)$\otimes$U(1)
for strong and electroweak interactions, has been remarkably successful
in particle physics.
In this model, however, there are too many coupling constants.
A straightforward extension of this model is SU(5).
The critical drawback of this model is, however, the nonobservation of the
proton decay predicted in the model.
There have been many grand unification models
including SU(15).\cite{FL}
We have no profound reason to prefer one model to others.

SU(15) is a nonrenormalizable theory since there is no anomaly cancellation
in itself. 
One way to avoid the nonrenormalizability is to introduce mirror
fermion. Such introduction of mirror fermion is not favorable but not
excluded.\cite{FL}
Therefore, it is worth reviewing the possible processes arising from SU(15).

In this letter, we investigate the possibility to discover the like-sign
dilepton production,
$e^-P\ra\bar\nu_e l^-l^- +anything$,
in SU(15) in the future $e^- P$ collider.
The dileptons are produced via the exchange of new gauge bosons
and they can be electrons, muons, or taus.
Here, we consider two muons to compare our result to
that of Agrawal {\it et al.}\cite{AFN}

We used the HELAS program\cite{HELAS}
in the numerical calculation of the cross section
of the process and the VEGAS program\cite{VEGAS}
for the multi-dimensional (10-dimensional
in this case) integrations stemmed from the phase spaces of
the kinematic variables of the particles, $\bar\nu_e$,
$\mu^-$'s and the EHLQ2 nucleon structure function set\cite{PDF}
for the $d$ quark structure function.
The HELAS program is designed mainly to provide numerical means for the
calculation of cross sections in the standard model.
However, it is easy to implement new interactions from extended standard
models like SU(15).

The process that we considered is $e^-P\ra\bar\nu_e \mu^-\mu^- +anything$
through the exchange of heavy gauge bosons $X^-$ or $X^{--}$.
We make use of the dilepton interaction Lagrangian given in the paper of
Frampton {\it et al.}\cite{FN}

\ba
{\cal L}_X =&&
   {g_{3l} \over \sqrt 2} X^{++}_\mu e^{T} C \gamma^\mu \gamma_5 e
  -{g_{3l} \over \sqrt 2} X^{--}_\mu \bar e \gamma^\mu \gamma_5 C \bar e^T 
     \nonumber\\
 &+&{g_{3l} \over \sqrt 2} X^+_\mu e^{T} C \gamma^\mu
   \Bigg( {1 - \gamma_5 \over 2} \Bigg) \nu_e
  +{g_{3l} \over \sqrt 2} X^-_\mu \bar \nu_e \gamma^\mu
   \Bigg( {1 - \gamma_5 \over 2} \Bigg) C \bar e^T.
\label{lagr}
\ea
Where, $X^-$($X^+$) and $X^{--}$($X^{++}$) are the massive lepton gauge bosons
which carry two
lepton numbers and result from the SU(3)$_l$ breaking.
The coupling constant $g_{3l}$ is given approximately by
$1.19e$.  Notice that vector couplings are missing for the currents
with the doubly charged gauge bosons due to Fermi statistics.

For the process, $e^-P\ra\bar\nu_e \mu^-\mu^- +anything$,
in $e^-P$ collider, we are interested in the subprocess,
$e d \ra \bar\nu_e \mu^-\mu^- u$,
through the exchange of gauge bosons, $X$'s.
Where, $d$ quark in the process comes as a parton inside the proton.
This process produces two muons and an electron-neutrino, $\bar\nu_e$,
in the final state while the only incoming lepton is the electron.
Therefore, although the total lepton numbers are conserved,
lepton numbers in each generation are violated.
Since the standard model is free from this violation, this process
produces a clean signal without such background processes in the standard model.
The relevant Feynman graphs for the process are shown in figs. 1-4.

\begin{figure}[h]
\vspace*{13pt}
\centerline{\psfig{file=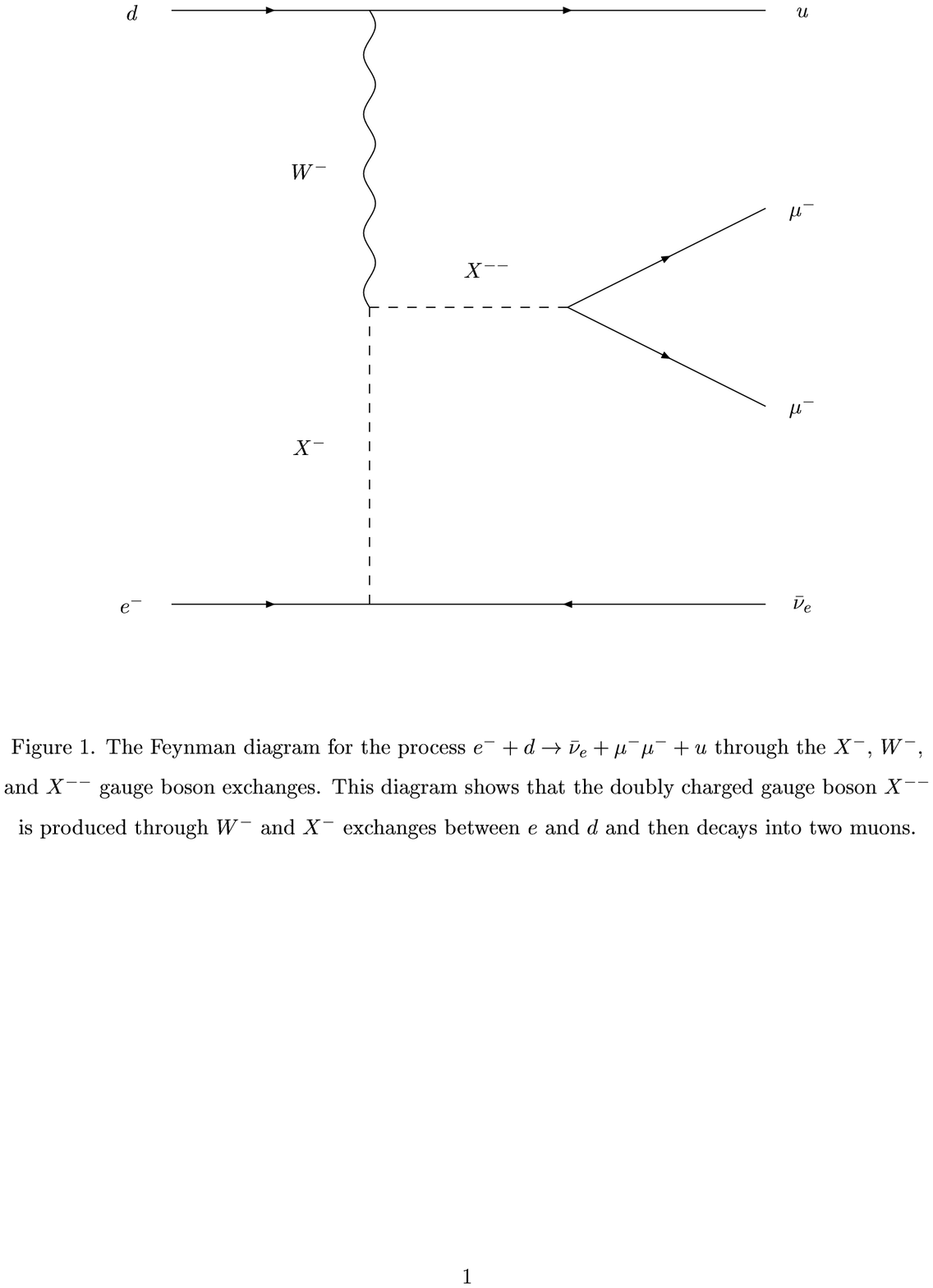}} 
\vspace*{13pt}
\caption{
The Feynman diagram for the process
$e^- +d \ra \bar\nu_e +\mu^-\mu^- +u$
through the $X^-$, $W^-$, and $X^{--}$ gauge boson exchanges.
This diagram shows that the doubly charged gauge boson $X^{--}$ is produced
through
$W^-$ and $X^-$ exchanges between $e$ and $d$ and then decays into two muons.
}
\label{fig1}
\end{figure}

\begin{figure}[h]
\vspace*{13pt}
\centerline{\psfig{file=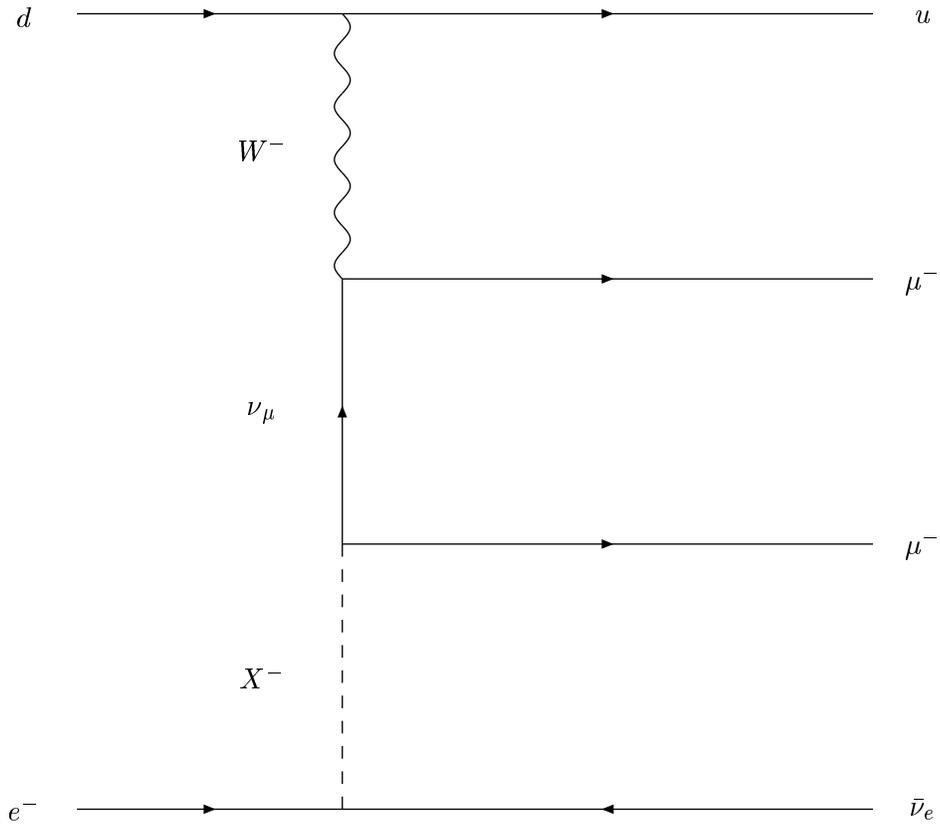}} 
\vspace*{13pt}
\caption{
The Feynman diagram for the process
$e^- +d \ra \bar\nu_e +\mu^-\mu^- +u$
through the $X^-$ gauge boson exchange.
This shows that the dileptons are produced through $W^-$ and $X^-$ with
the exchange of a neutrino.
There are two graphs for this case. This is one of them.
}
\label{fig2}
\end{figure}

\begin{figure}[h]
\vspace*{13pt}
\centerline{\psfig{file=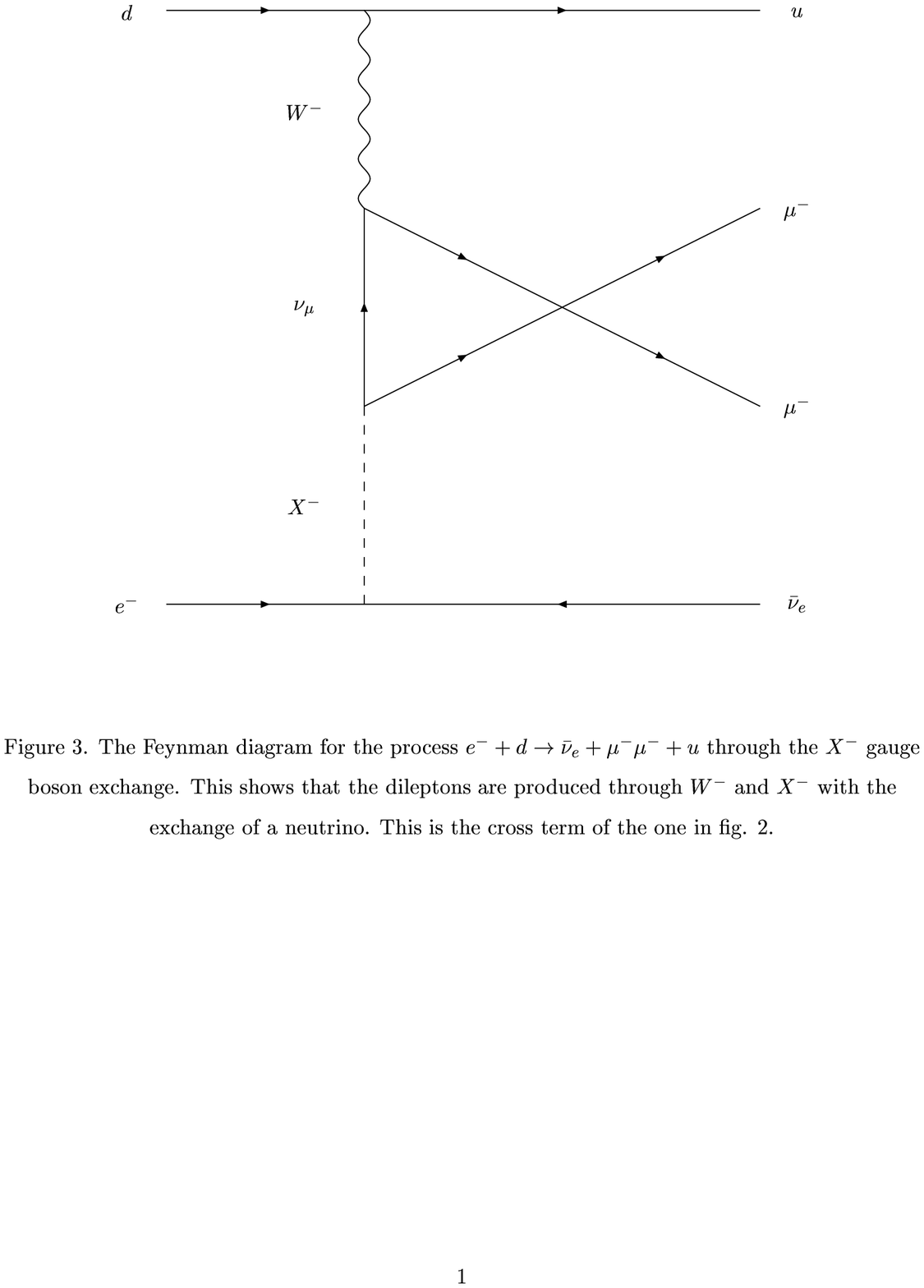}} 
\vspace*{13pt}
\caption{
The Feynman diagram for the process
$e^- +d \ra \bar\nu_e +\mu^-\mu^- +u$
through the $X^-$ gauge boson exchange.
This shows that the dileptons are produced through $W^-$ and $X^-$ with
the exchange of a neutrino. This is the cross term of the one in fig. 2.
}
\label{fig3}
\end{figure}

\begin{figure}[h]
\vspace*{13pt}
\centerline{\psfig{file=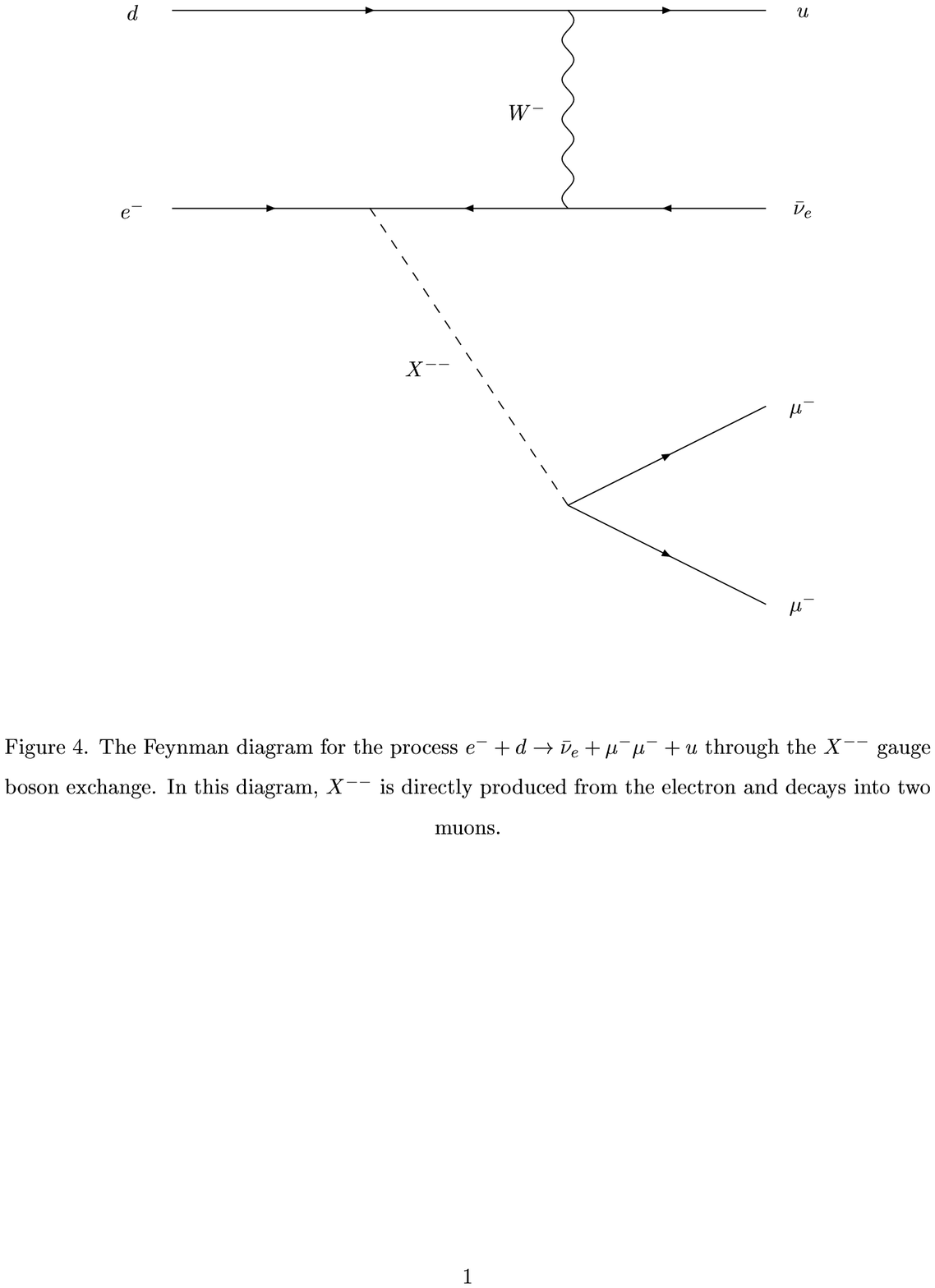}} 
\vspace*{13pt}
\caption{
The Feynman diagram for the process
$e^- +d \ra \bar\nu_e +\mu^-\mu^- +u$
through the $X^{--}$ gauge boson exchange.
In this diagram,
$X^{--}$ is directly produced from the electron and decays into two muons.}
\label{fig4}
\end{figure}

The new gauge bosons can be produced from the incoming lepton.
Then, they produce two muons directly (as shown in fig. 4) or
interact with $W$ boson to produce two muons (as shown in figs. 2 and 3) or
interact with $W$ boson to produce another doubly charged gauge boson
which then turns into two muons (as shown in fig. 1).
The corresponding amplitudes are obtained from the Lagrangian
given in eq. (\ref{lagr}),

\ba
M(1) &=&
\bar u_u \Bigg[ -{ig \over \sqrt 2 } \gamma_\mu
  \Big( {1-\gamma_5 \over 2} \Big) U_{ud} \Bigg] u_d
\Bigg[ {-i( g^{\mu\nu_1} - k_W^\mu k_W^{\nu_1}/M_W^2 )
  \over k_W^2 - M_W^2} \Bigg]
\nonumber\\
&&\Big( -{ig_{3l} \over \sqrt 2} \Big) \bar u_{\mu_1} \gamma_\rho \gamma_5
 C \bar u_{\mu_2}^T
\Bigg[ {-i( g^{\rho \nu_3} - k_{X^{--}}^\rho k_{X^{--}}^{\nu_3}/M_{X^{--}}^2 )
        \over k_{X^{--}}^2 - M_{X^{--}}^2} \Bigg]
\nonumber\\
&&\Big( -{ig \over \sqrt 2 } \Big)
\Big[ (k_{X^-} + k_{X^{--}})^{\nu_1} g^{\nu_2\nu_3}
     - k_{X^-}^{\nu_3} g^{\nu_1\nu_2} - k_{X^{--}}^{\nu_2} g^{\nu_3\nu_1} \Big]
\nonumber\\
&&\Big( -{ig_{3l} \over \sqrt 2 } \Big)
v_{\nu_e}^T C \Big( {1-\gamma_5 \over 2} \Big) \gamma_\sigma u_e
\Bigg[ {-i( g^{\nu_2\sigma} - k_{X^-}^{\nu_2} k_{X^-}^\sigma/M_{X^-}^2 )
  \over k_{X^-}^2 - M_{X^-}^2} \Bigg],
\nonumber\\
M(2) &=&
\bar u_u \Bigg[ -{ig \over \sqrt 2 } \gamma_\mu
  \Big( {1-\gamma_5 \over 2} \Big) U_{ud} \Bigg] u_d
\Bigg[ {-i( g^{\mu\nu} - k_W^\mu k_W^\nu/M_W^2 )
  \over k_W^2 - M_W^2} \Bigg]
\nonumber\\
&&\Big( -{ig \over \sqrt 2} \Big) \bar u_{\mu_1} \gamma_\nu
\Big( {1-\gamma_5 \over 2} \Big)
\Bigg[ {i\over\rlap/ k_{\nu_\mu}-m_{\nu_\mu}}\Bigg]
\nonumber\\
&&\Big( -{ig_{3l} \over \sqrt 2} \Big) \gamma_\rho
 \Big( {1-\gamma_5 \over 2} \Big) C \bar u_{\mu_2}^T
\Bigg[ {-i( g^{\rho\sigma} - k_{X^-}^\rho k_{X^-}^{\sigma}/M_{X^-}^2 )
        \over k_{X^-}^2 - M_{X^-}^2} \Bigg]
\nonumber\\
&&\Big( -{ig_{3l} \over \sqrt 2 } \Big)
v_{\nu_e}^T C \Big( {1-\gamma_5 \over 2} \Big) \gamma_\sigma u_e,
\nonumber\\
M(3) &=& (\mu_1 \leftrightarrow\mu_2),
\nonumber\\
M(4) &=&
\bar u_u \Bigg[ -{ig \over \sqrt 2 } \gamma_\mu
  \Big( {1-\gamma_5 \over 2} \Big) U_{ud} \Bigg] u_d
\Bigg[ {-i( g^{\mu\nu} - k_W^\mu k_W^\nu/M_W^2 )
  \over k_W^2 - M_W^2} \Bigg]
\nonumber\\
&&\Big( -{ig_{3l} \over \sqrt 2} \Big) \bar u_{\mu_1} \gamma_\rho \gamma_5
  C \bar u_{\mu_2}^T
\Bigg[ {-i( g^{\rho\sigma} - k_{X^{--}}^\rho k_{X^{--}}^{\sigma}/M_{X^{--}}^2 )
        \over k_{X^{--}}^2 - M_{X^{--}}^2} \Bigg]
\nonumber\\
&&\Big( -{ig \over \sqrt 2} \Big) 
v_{\nu_e}^T C \Big( {1-\gamma_5 \over 2} \Big) \gamma_\nu
\Big( {i \over \rlap/ p_e - m_e} \Big)
\Big( -{ig_{3l} \over \sqrt 2} \Big) \gamma_\sigma \gamma_5 u_e.
\label{amp}
\ea
Where, $\mu_1$ and $\mu_2$ indicate two muons.
The vertex, $WX^- X^{--}$, is similar to triple gauge boson vertices
like $WWW$ or $ggg$ in the standard model.
$U_{ud}$ is the Cabbibo-Kobayashi-Maskawa (CKM) matrix element
for $u$ and $d$ quarks.
Here, we have used unitary gauge for the gauge bosons to
make use of HELAS subroutines which are written in unitary gauge
when the bosons are massive.
Only a small modification such as the change in couplings is needed
to apply HELAS subroutines to this extended standard model.
The total amplitude for the process
$e^- d \ra \bar\nu_e \mu^-\mu^- u$
is the sum of all the amplitudes in eq. (\ref{amp}),
\ba
M=M(1)+M(2)+M(3)+M(4).
\ea

We can simplify the differential cross section for the process by
writing the four body phase space as a product of a few two body phase spaces.
\ba
d\hat\sigma&=&{\overline{|M|^2} \over 2(2\pi)^8 \lambda^{1/2}(\hat s,m_e,m_d)}
d_4 (PS),
\nonumber\\
d_4 (PS)&=&d_2(e+d \ra A+B) dm_A^2 dm_B^2 d_2(A \ra \bar \nu_e u)
    d_2(B \ra \mu^- \mu^-).
\label{crosec}
\ea
Where, $d\hat\sigma$ is the differential cross section for a parton (quark)
$q$ from the proton to be scattered by the electron.
$\hat s=xs$ is the square of the center of mass energy of the
electron and the parton from the proton with the total center of mass energy
$\sqrt s$ of the system of the incoming electron and the incoming proton.
$x$ is the standard Bjorken variable and thus
the $d$ quark carries the fractional momentum of the proton,
$x$ times the proton momentum.
The four body phase space $d_4 (PS)$ can be obtained from the process
$e^- d \ra \bar\nu_e \mu^-\mu^-u$
via the process $e^-+d\ra A+B$ and then
$A\ra \bar\nu_e +u$ and $B\ra \mu^-+\mu^-$.
The two body phase spaces are given as
\ba
d_2(e+d \ra A+B) &=& {1\over 8 \hat s}
 \lambda^{1/2}(\hat s, m_e^2,m_d^2) d \Omega_B^*, \nonumber\\
d_2(A \ra \bar \nu_e u) &=& {1\over 8 m_A^2}
 \lambda^{1/2}(m_A^2, m_u,m_\nu) d \Omega_u^*, \nonumber\\
d_2(B \ra \mu^- \mu^-) &=& {1\over 8 m_B^2}
 \lambda^{1/2}(m_B^2, m_{\mu^-}^2,m_{\mu^-}^2) d \Omega_\mu^*.
\label{d2ps}
\ea
$m_A$ is the invariant mass of the system of the $u$ quark and the antineutrino
and $m_B$ is the invariant mass of the system of two muons.
Angular variables are defined in their center of mass frames.
$d\Omega^*_B$ is defined in the center of mass frame of $A$ and $B$.
Similarly, $d\Omega^*_u$ is defined in the center of mass frame of
$u$ and $\bar\nu_e$ and $d\Omega^*_\mu$ in that of two muons, $\mu^-$ and
$\mu^-$.

We can calculate the total cross section by summing over all the partons
and integrating over $x$.
\be
\sigma = \int_0^1 dx\sum_q F_q(x,Q^2) \hat \sigma (\sqrt {\hat s}=xs,M_X).
\label{totcrosec}
\ee
Here, instead of summing over all flavors,
we need only $d$ quark distribution function
to produce two like-sign muons and an antineutrino in the final state.
Therefore, the summation of the distribution functions over all the flavors
can be simply replaced by the $d$ quark distribution function,
$\sum_q F_q(x,Q^2) \ra d(x,Q^2)$.
We use this $d$ quark structure function from
the EHLQ2 nucleon structure function set in the PDF-library routines
in the Phenomenology Institute at the Univesity of Wisconsin.
Taking different nucleon structure function sets such as EHLQ1 or KMRSB
does not make any physically significant difference.

The numerical value of the cross section of the process is obtained by
using HELAS and VEGAS programs.
The VEGAS adaptive monte carlo integration subroutines were used
for the multi-dimensional integrations arising from the kinematics of
the multi-particles involved in the process.
In the VEGAS program, the `FUNCTION' subprogram for the cross section
has to be defined in connection with the HELAS program.
This FUNCTION subprogram represents the cross section and thus the HELAS
program plays the central role in FUNCTION.

Before starting HELAS, we have to prepare a few subroutines.
For the coupling which involve in the interactions of the process,
the relevant coupling subroutines (COUP1X and COUP2X) should be determined.
To find the momenta of the particles in the center of mass frame,
we use the momentum subroutine (MOM2CX) along with
the subroutines (MOMNTX, BOOSTX) related to the boost factors.
Now, we draw the Feynman diagrams of this process and then
apply HELAS Feynman rules.
This is the central part of the application of HELAS.
After setting up all these, we are ready to use HELAS and follow the steps
below.

First, we find the wave functions (IXXXXX, OXXXXX) of the external particles
by identifying flowing in and flowing out particles (all are fermions
in this case).
Second, we find the vertex subroutines
(JIOXXX, FVOXXX, VVVWXX, IOVXXX)
for the corresponding vertices in the Feynman diagrams.
Third, we calculate the total amplitude with the phase space factors.
Fourth, we multiply EHLQ2 $u$ quark structure function (EHLQ2PDF).
Agrawal {\it et al.} used EHLQ1 for the $u$ quark structure function.\cite{AFN}
To obtain the numerical result for the cross section,
we have to integrate over all relevant kinematic variables by running
VEGAS with the above FUNCTION subprogram.

In $e^- P$ collider at HERA, electrons of the energy 30GeV 
collide on protons of the energy 820GeV in producing the center of mass energy,
$\sqrt{s}=314$GeV.
We chose a few values of the mass of the gauge boson, $M_X\approx$100GeV,
200GeV, or 300GeV.
We also used the constituent quark masses, $m_u =0.3$GeV and $m_d =0.3$GeV.
All other standard values were used in the numerical calculation such as
$M_W=80.22$GeV and $\sin\theta_W=0.2325$GeV, etc.
With all the values considered, we obtained the following results
for the cross section (see Table 1).

The cross section of the process is $8\times10^{-2}$fb for $M_X=100$GeV,
$3\times10^{-4}$fb for $M_X=200$GeV, and $3.5\times10^{-6}$fb for $M_X=300$GeV.
These results show that the cross section decreases very sharply as
the gauge boson $X$ increases.
Since the cross section is very small,
it is quite difficult to observe the process in the foreseeable future.
The result is also an order of 2 smaller than the one
in the process considered by Agrawal, Frampton, and Ng.\cite{AFN}
There, they considered the process, $e^- q\ra e^+2\mu^-q$.
In the process, there is no quark flavor changes and
there is a positron instead of antineutrino in the final state.
The behavior of the sharp declination of the cross section
as the increment of $M_X$
is similar to both our process and theirs.
Although the cross section of our process is much smaller than
that of theirs, our process does not contain well measurable particle,
$e^+$, in the final state with different kinematics.
In our process, there is the missing momentum due to neutrinos without
electron or positron in the final state.
Therefore, we hope to be able to isolate our process from theirs
in the collider of next generation.
Unfortunately, however, considering the current luminocity
$\sim10^{31}$ cm$^{-2}$s$^{-1}$ of HERA
we expect that the cross section of this process is too small
to be observed in near future.

Frampton and Ng\cite{FN} showed that the lower bound of the gauge boson mass,
$M_X$, is greater than 120GeV from the experimental angular distribution
in Bhabha scattering, $e^+e^-\ra e^+e^-$.
Carlson and Frampton\cite{CN} concluded $M_X>$230GeV from the precision data
of the angular distribution of electrons in polarized muon decay.
Since the value of the coupling, $g_{3l}\approx1.19e$ is not
the unavoidable choice, the above mass lower bound is not strict.
And thus Agrawal {\it et al.}\cite{AFN} calculated the cross section
of the process,
$e^- q\ra e^+2\mu^-q$,
with the broad range of $M_X$, 100GeV $<M_X<$ 1TeV.
We also calculated the cross section of the process,
$e^- P\ra \bar\nu_e \mu^- \mu^- +anything$,
starting with $M_X=$100GeV to compare
our result to that of Agrawal {\it et al.}\cite{AFN}.
However, the lower bound of $M_X$ should be well above 100GeV.
Another good way to find the mass limit of the dilepton gauge boson may be
to look into the precision electroweak data at the $Z$ peak.
The loop calculation with these dilepton gauge bosons should modify
the radiative correction to $Z$ or $W$ boson propagators.
The comparision of the result with the current data may set a better
lower limit on $M_X$.
We have to keep in mind that SU(15) is not a renormalizable theory
and some modifications have to be made in the application of the
HELAS program in this calculation.

\begin{table}[htbp]
\tcaption{
The cross section for $e^- +d\ra\bar\nu+u+\mu^-+\mu^-$ with various
value of $M_X$
\label{table1}
}
\centerline{\footnotesize\smalllineskip
\begin{tabular}{c c c}\\
\hline
$M_X$(GeV)& cross section in (fb)&
cross section of AFN\cite{AFN} in (pb) \\
\hline
100& $\sigma\sim8.0\times10^{-2}$& $\sigma\sim10^{-2}$ \\
200& $\sigma\sim3.0\times10^{-4}$& $\sigma\sim10^{-4}$ \\
300& $\sigma\sim3.5\times10^{-6}$& \\
\hline\\
\end{tabular}}
\end{table}

\nonumsection{Acknowledgments}
\noindent
We would like to thank Prof. P. Ko and Prof. M. Voloshin for useful discussions
and the phenomenology group at the University of Wisconsin for providing us
the PDF-library functions.
We also thank YVRC at Yonsei University for the hospitality.

\nonumsection{References}

\end{document}